# Bismuth-content dependent of the polarized Raman spectra of the *InPBi* alloys

G. N. Wei[1], Q. H. Tan[1], Q. Feng[1], D. Xing[1], W. G. Luo[1], K. Wang[2], L. Y. Zhang[2], S. M. Wang[2] and K. Y. Wang[1, a)]

[1] SKLSM, Institute of Semiconductors, CAS, P. O. Box 912, 100083, Beijing, P. R. China

[2] State Key Laboratory of Functional Materials for Informatics, Shanghai Institute of Microsystem and Information Technology, CAS, 865 Changning Road, Shanghai 200050, China

**Abstract:**

We have systematically investigated the optical properties of the $InP_{1-x}Bi_x$ ternary alloys with $0 \leq x \leq 2.46\%$, using high resolution polarized Raman scattering measurement. Both *InP*-like and *InBi*-like optical vibration modes (LO) were identified in all the samples, suggesting most of the *Bi*-atoms are incorporated into the lattice sites to substitute *P*-atoms. We found the intensity of the *InBi*-like Raman modes increase exponentially as Bi-content increasing. Linearly red-shift of the *InP*-like longitudinal optical vibration modes was observed to be 1.1 cm$^{-1}$ of percent Bi, while that of the *InP*-like optical vibration overtones (2LO) were nearly doubled. In addition, through comparing the difference between the $Z(X,X)\bar{Z}$ and $Z(X,Y)\bar{Z}$ Raman spectra, Longitudinal-Optical-Plasmon-Coupled (LOPC) modes are identified in all the samples, and their intensities are found to be proportional to the electron concentrations.

## I. INTRODUCTION

The temperature-insensitive semiconductor laser is one of the future key devices in the field of optical communications.[1] Bismuth-diluted alloys are expected to have temperature-insensitive band gaps[1-3], and therefore are the promising candidates for the application in the optoelectronic devices. Berding et al. theoretically predicted that *InPBi* is the best potential candidate for mid- and far-infrared (IR) optoelectronics

---

[a)] E-Mail: kywang@semi.ac.cn

applications among *InSbBi*, *InAsBi* and *InPBi*.[3] Thus isoelectronic *Bi* doping in *InP* has become an important subject in the search for a new semiconducting material.[1] However, the *InPBi* alloy was the most difficult to mix, and not until recently, have the $InP_{1-x}Bi_x$ materials been successfully grown by the MBE technique[1,2].

Raman scattering study is a very powerful technique to study the crystal quality and the related vibration properties[4]. The vibration properties of the *InPBi* materials have been studied using un-polarized Raman scattering measurement, where the *InBi*-like optical vibration modes are identified with Raman frequencies of 148 and 170 cm$^{-1}$.[5,6] Yet in the previous work those *Bi*-induced modes are not subtracted from the background signals provided by the *InP*-like acoustical vibration modes. And the Longitudinal-optical-Plasmon-coupled (LOPC) modes have not been identified, nor did the shift of the *InP*-like optical modes. Different to the previous work, here we use polarized micro-Raman scattering system with much higher resolution to further investigate the *Bi*-content dependent vibration properties of the $InP_{1-x}Bi_x$ alloys with $0 \leq x \leq 2.46\%$. To compare the Raman spectra under different polarized conditions, we are able to assign the *InBi*-like Raman features at about 148 and 170 cm$^{-1}$ to be out-of-plane and in-plane vibration modes, respectively. Intensities of the *InBi*-like optical vibration modes were found to increase exponentially with increasing the Bi-content. Linearly red-shift of the *InP*-like longitudinal optical vibration modes was observed with 1.1 cm$^{-1}$/*Bi*%, while that of *InP*-like optical vibration overtones (2LO) were nearly doubled. In addition, the appearance of the Longitudinal-Optical-Plasmon-Coupled (LOPC) mode has been verified by comparing two polarized Raman spectra, and the *Bi* doping content dependence of the LOPC intensity and the free-electron concentrations have the same trend.

## II. EXPERIMENT

### A. Sample details

The 400 nm thick $InP_{1-x}Bi_x$ films, with $0 \leq x \leq 2.46\%$ were grown on (100) semi-insulating *InP* substrates by V90 gas source molecular beam epitaxy (GSMBE) at fixed growth temperature 193℃. *Bi* compositions were determined by Rutherford backscattering spectrometry (RBS) with 2.275 MeV $^4$He$^{2+}$ ions, with details published elsewhere.[1,2] The epilayers were cut into a size of a square of cm for Raman scattering measurements, with mirror like surface.

**B. Raman setup**

Raman measurements were performed on a Renishaw inVia plus at room temperature. The 1.97 eV line of an Argon-ion laser with a power of 0.5 mW, was used for excitation. The scattered light was analyzed by a triple monochromator (Dilor XY) equipped with aliquid-nitrogen-cooled CCD array. Using a high precision microscope, the laser spot size can be focused on around 1 μm in diameter, which enables us to investigate the spatial homogeneity of the samples. A 2400 g/mm Diffraction grating (with ~0.2 cm$^{-1}$) was used. Correct instrument calibration was verified by checking the position of the Si band at $\pm$ 520.7 cm$^{-1}$. The widely used Porto notations $Z(X,X)\bar{Z}$ and $Z(X,Y)\bar{Z}$ have been used for the designation of crystal and polarization directions in this work. In the polarized Raman measurements, for $Z(X,X)\bar{Z}$ configuration, a polarization analyzer was placed right after the edge filter and the polarization direction was parallel to the polarization of incident laser beam. The $Z(X,Y)\bar{Z}$ configuration was conducted by placing a half-wave plate before the analyzer in $Z(X,X)\bar{Z}$ configuration. Because the thicknesses of *InPBi* film are large compared to the optical absorption depth, the signal contribution from the substrate is negligible.

## III. RESULTS AND DISCUSSION

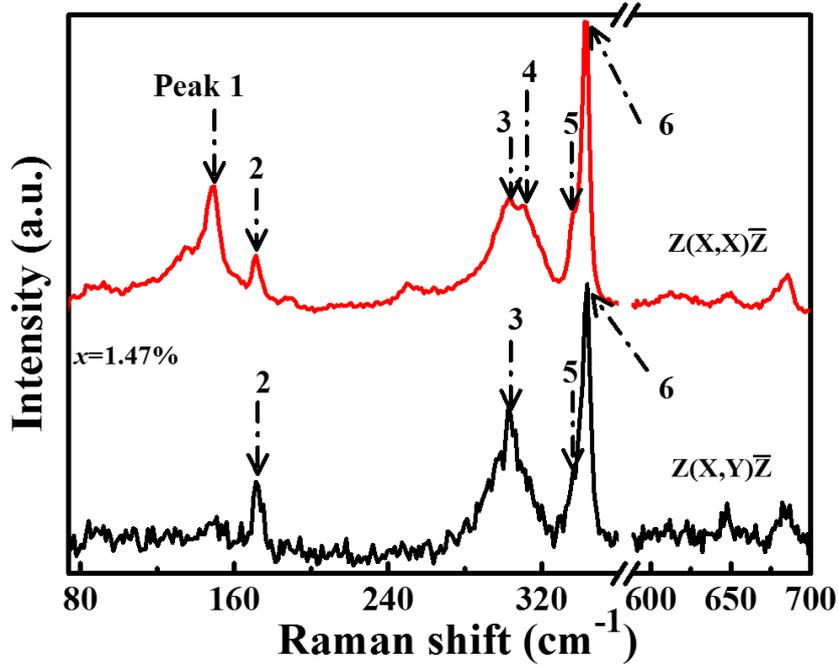

*FIG. 1. Raman spectra for InP$_{1-x}$Bi$_x$ alloys with x=1.47%, under $Z(X,X)\bar{Z}$ (red line) and $Z(X,Y)\bar{Z}$ (black line) configurations.*

Polarized Raman scattering studies allow us to further confirm the symmetry of the main Raman features. Taking the *InP$_{1-x}$Bi$_x$* alloys with *x*=1.47% for example, Figure 1(a) shows the Raman features obtained under both the $Z(X,X)\bar{Z}$ (red line) and $Z(X,Y)\bar{Z}$ (black line) configurations. For Raman spectra under $Z(X,X)\bar{Z}$ configuration, there are mainly six Raman features at 148, 170, 303, 311, 336, and 344 cm$^{-1}$ marked as 1~6 between 80 and 350 cm$^{-1}$. Raman peaks 2, 3, 5, and 6 also existed in Raman spectra under $Z(X,Y)\bar{Z}$ configuration, while peaks 1 and 4 were missing. Raman peaks 3~6 can be assigned as the *InP*-like Raman optical modes[7,8]. Accordingly, Raman peaks 3, and 4 can be assigned as *InP*-like transverse optical vibration modes at Γ point TO(Γ) and X point TO(X) respectively, while for Raman peaks 5 and 6, the InP-like longitudinal optical vibration modes at L or X point LO(L or X) and at Γ point LO(Γ), respectively[6]. The sharp and strong *InP*-like Raman longitudinal optical vibration mode LO(Γ) (peak 6), indicating the existence of long range order of the *InP* structure in *InPBi* epilayer[7]. *InP* semiconductors with typical III-V group zinc-blende structure, show *Td* site symmetry. Unpolarized Raman spectra recorded from the *InP*

(100) surfaces, shows the contributions of the three irreducible components $\Gamma_1$, $\Gamma_{12}$, and $\Gamma_{15}$.[7] Thus the Raman spectrums in $Z(X,X)\bar{Z}$ configuration, containing $\Gamma_1 \oplus 4\Gamma_{12}$ symmetry, show the in-plane vibrational modes. While those in $Z(X,Y)\bar{Z}$ configuration, with $\Gamma_{15}$ symmetry, show the out-of-plane vibration modes[6]. This well explains why the Raman peaks 3, 5, and 6 should be allowed in the $Z(X,X)\bar{Z}$ and $Z(X,Y)\bar{Z}$ Raman scattering geometries, while Raman peak 4 went missing in the $Z(X,Y)\bar{Z}$ configuration.

The Raman peaks 1 and 2, belonging to the *InBi*-like optical vibration modes[5], can be assigned as Raman optical modes TO and LO. The *InBi* mode in $InP_{1-x}Bi_x$ should originate from the substitutional Bi atoms at the P site, resulting in the same Raman selection rules for both *InBi* and *InP* modes. Therefore, the Raman selection rules in the $Z(X,X)\bar{Z}$ and $Z(X,Y)\bar{Z}$ scattering geometries for the *InBi* mode in $InP_{1-x}Bi_x$ alloys[9]. The optical phonon frequencies of *InBi* mode can be estimated from $\omega_{InBi} = \omega_{InP} \cdot \sqrt{\mu_{InP}/\mu_{InBi}}$, where $\mu_{InP}$ and $\mu_{InBi}$ are the reduced masses of *In–P* and *In–Bi*, respectively.[10] Considering that the TO (peak4) and LO (peak 6) phonon frequencies of *InP*, the *InBi* mode is expected to be near the spectral range between 150 and 200 cm$^{-1}$. This can also well explain why both the peak 1 and 4 went missing in the $Z(X,Y)\bar{Z}$ Raman scattering geometries.

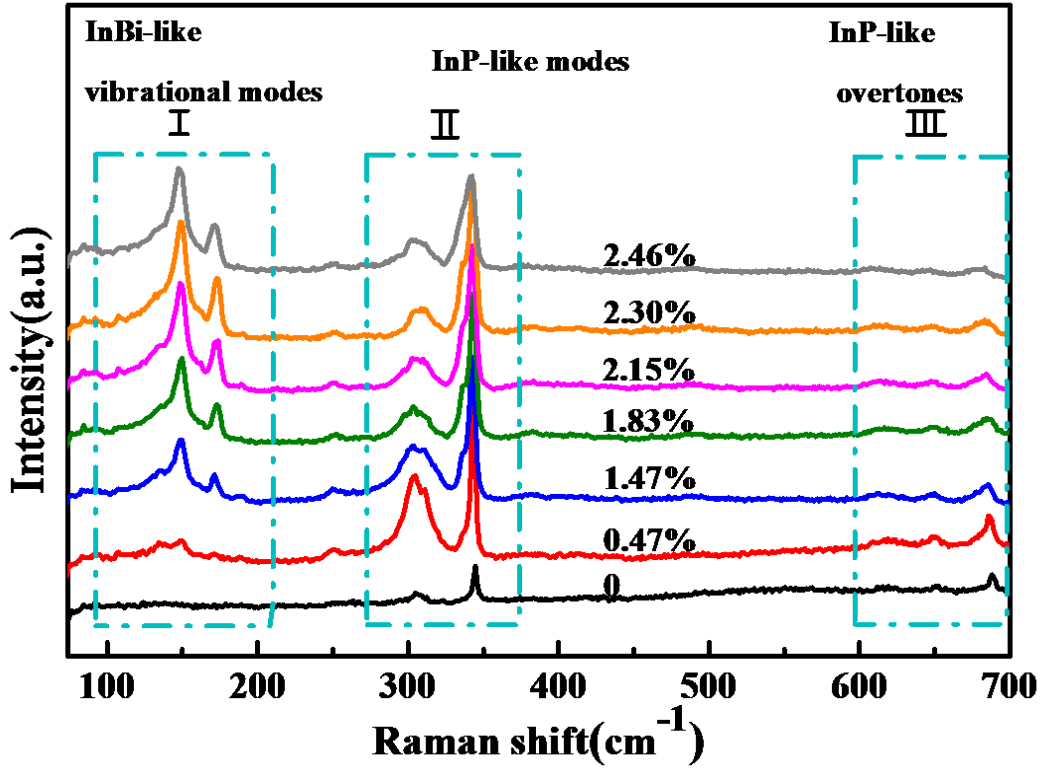

*FIG. 2. Raman spectra under $Z(X,X)\bar{Z}$ configurations for $InP_{1-x}Bi_x$ samples, with $0 \leq x \leq 2.46\%$, on the range from 75 to 700 $cm^{-1}$. InBi-like optical vibration modes are observed at low-frequency region (marked I). InP-like optical vibration modes are observed at medium frequency region (marked II), while InP-like optical vibration overtones are observed at high frequency region (marked III).*

With many more Raman features appeared in Raman spectra under $Z(X,X)\bar{Z}$ configuration, we present in figure 3 the *Bi*-content dependent $Z(X,X)\bar{Z}$ polarized Raman spectra for the series of $InP_{1-x}Bi_x$ alloys with $0 \leq x \leq 2.46\%$. In order to better analyze the spectra, the Raman features are marked as three different regions, I (low-frequency region), II (medium-frequency region) and III (high-frequency region), respectively. In the following, we will discuss the *Bi*-composition dependent Raman features in details, one section by one section.

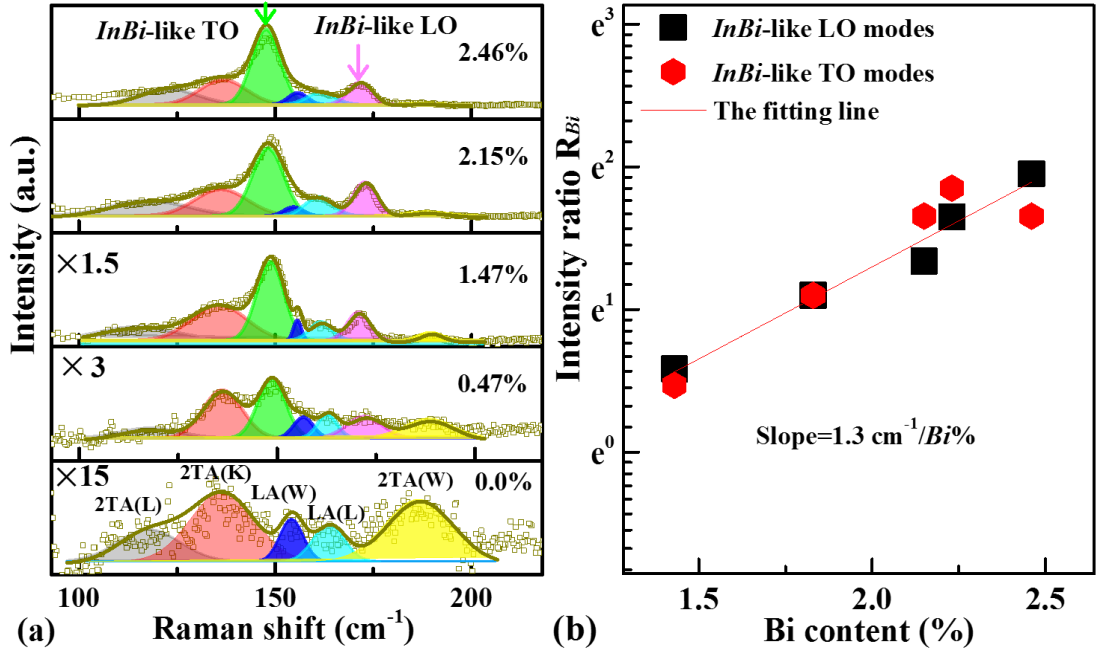

FIG. 3. (a) InBi-like vibration modes TO (represented by the red section) and LO (purple section), together with InP-like disorder-activated acoustical modes were used in the multi-peak fitting on the low frequency range from 93 to 218 cm$^{-1}$. The intensity of the Raman spectra were multiplied by 15, 3 and 1.5 for $x$ =0, 0.47% and 1.47% samples, respectively. (b)The Bi content dependent the TO (red solid) and LO (black square) intensity ratio respect to the InP$_{1-x}$Bi$_x$ sample with x=0.47%, with $R_{Bi}$=[I(x)-I(0.47%)]/I(0.47%), where the line is the fitted results.

The low frequency region Raman spectra for the *InP$_{1-x}$Bi$_x$* alloys with $0 \leq x \leq 2.46\%$ were shown in figure 3a. In order to see the weak peaks clearly, the intensity of the Raman spectra were multiplied by 15, 3 and 1.5 for $x$ =0, 0.47% and 1.47% samples, respectively. High resolution techniques, allow us to identify those weaker background signals as Raman features at about 117, 135, 154, 162, and 188 cm$^{-1}$, which also appeared in the *InP* reference samples. Those five weaker Raman features are assigned to the *InP*-like transverse or longitudinal acoustical modes and their overtones, 2LA(L), 2TA(K), LA(W), LA(L), 2TA(W).[7,8] With adding the *Bi* atoms into the *InP* crystal, two extra peaks at 148.5 and 171.5 cm$^{-1}$ were observed, where the first one is assigned to *InBi*-like TO mode and the second one to LO mode. The

Raman peak intensity for both the *InBi*-like TO and LO modes increases with increasing the Bi doping level, while the peak positions for these two modes are not sensitive to the Bi doping concentration. In order to characterize the Bi-content dependent intensity and peak-site behavior of the *InBi*-like Raman features, we have to rule out the background scattering features, where the peak positions for the background signals should be not sensitive to the variation of the Bi composition. The Raman spectra for all the Bi doped samples can be well fitted by considering the *InP* background signals and the *InBi* like modes, which are shown in Figs. 3(a). The intensity of the *InBi*-like TO and LO modes respect to the intensity of $InP_{1-x}B_x$ sample with $x=0.47\%$, $R_{Bi}=[I(x)-I(0.47\%)]/I(0.47\%)$, is shown in Figs. 3(b). The intensity for both the *InBi*-like TO and LO modes exponentially increases with increasing the *Bi* content, where the relation between $R_{Bi}$ and *Bi* content ($x$) is $\ln(R_{Bi}) \sim x$, with the fitting slope value as 1.3 cm$^{-1}$/*Bi*%. It is worth noting that *Bi* content dependent $\ln(R_{Bi})$ for both the *InBi*-like TO and LO modes falls into the same trend, indicating these two modes are closely linked with increasing the *Bi* doping.

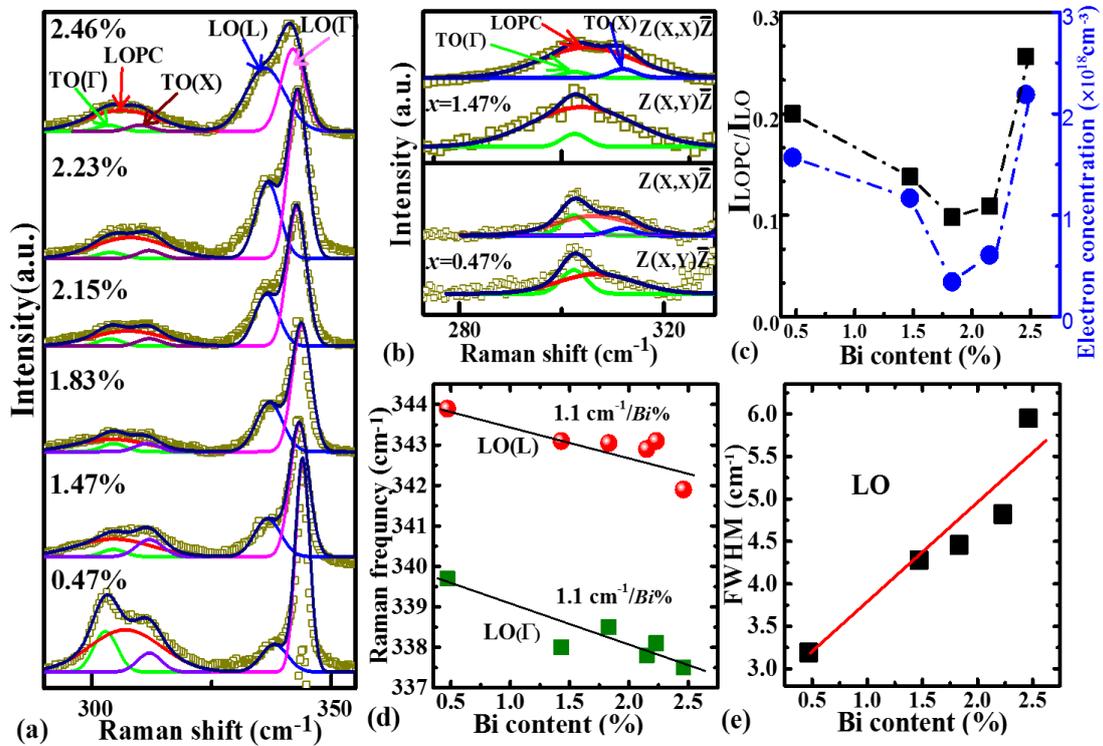

*FIG. 4. (a) The Raman spectra between 280 and 365 cm$^{-1}$ with different Bi doping*

*level, where the dark yellow dots are the experimental data. InP-like TO(L) (green line ), LOPC (red line), and TO(Γ) (purple line) modes are used to fit the spectra between 280 and 325cm$^{-1}$, so as LO(L) (blue line ) and LO(Γ) (magneta line ) for that between 325 and 365 cm$^{-1}$. Navy lines are the fitted curves. (b) Comparation the Raman spectra in the range of 270 to 330 cm$^{-1}$ between the $Z(X,X)\bar{Z}$ and $Z(X,Y)\bar{Z}$ polarized configurations, for samples with x=0.47% and 1.47%. (c) Bi content dependent intensity of the LOPC and the corresponding electron concentrations. (d) The Bi doping content dependent of the shift of the Raman frequencies of LO(L) and LO(Γ) modes. (e) The Bi composition dependent full width at half maximum (FWHM) of LO(Γ). The black dots represent the experiment data, while the red line guides the eye.*

Figure 4 presents the analysis of the *Bi* content dependent Raman spectra at the medium frequency region. Three different kinds of Raman features have to be introduced to fit the Raman spectra between 270 and 330 cm$^{-1}$, which are *InP*-like transverse optical vibration modes TO(Γ) (green peak), TO(X) (purple peak), and Longitudinal-Optical-Plasmon-Coupled mode LOPC (red peak), shown in Figs. 4(a). The reason is that if only Raman features of TO(Γ) and TO(L) exist in the Raman spectra under $Z(X,X)\bar{Z}$ configuration, then consequently there will be only TO(Γ) left in that under $Z(X,Y)\bar{Z}$ configuration. However, as shown in Figs. 4(b), the peak TO(L) does go missing under $Z(X,Y)\bar{Z}$ configuration, yet the width of the Raman curve stays as the same. So there must be another Raman features exist. Then, we will show that the most likely candidate is LOPC mode. In a polar semiconductor, the free-carrier Plasmon and the longitudinal-optical (LO) phonons are coupled by the interaction between the electric dipole moment due to the relative displacement of the ions and the electric field associated with the free carriers.[11] In that case, these modes will appear near the LO phonon frequency, also named as LO-phonon-plasmon coupled (LOPC) mode. Commonly, there are two coupled modes in heavily doped n-type *InP* materials[12], the upper ($L^+$) and lower ($L^-$) longitudinal branches of the coupled plasmon-LO phonon modes respectively. In *InP* with high electron concentrations

($n \sim 10^{18}$ cm$^{-3}$), the frequency of the upper branch is much higher than the LO phonon frequency, while that of the lower branch is closely equal to the TO phonon frequency[11]. Given the fact that the obtained electron concentrations of the *InPBi* materials are on the range of $10^{18}$ cm$^{-3}$, [13] thus the LOPC mode centered at about 303 cm$^{-1}$ was introduced in the fitting procession. After considering these three modes, the Raman spectra for different samples can be well fitted. For Raman spectra on the range between 330 and 350 cm$^{-1}$, two kinds of Raman features, which are the *InP*-like LO($\Gamma$) and LO(L) modes are introduced in the fitting progress, shown in Figs. 4(a).

Because the intensity of the LOPC mode is strongly dependent on the carrier densities, LOPC mode has been often used as a nondestructive probe to investigate the relative doping level of the n type semiconductors[14,15]. The intensity ratio $I_{LOPC}/I_{LO}$ was shown in Figs. 4(c). The ratio of $I_{LOPC}/I_{LO}$ does not vary linearly with Bi doping concentration, but has similar trend with the electron densities obtained from Hall Effect measurements, which proves that indeed the $I_{LOPC}$ can be used as a rapid and sensitive method to determine the relative free carrier density in n-type semiconductors.

Red shift of both the LO($\Gamma$) and LO(L) modes linearly increases with increasing the Bi doping level, which is shown in Figs. 4(d). The red shift of the Raman frequency for both the *InP*-like LO($\Gamma$) and LO(L) modes is to be 1.1 cm$^{-1}$/ *Bi*%. This may lay in the enlargement of lattice constant, which vary linearly from 5.86 to 5.88Å[3], as *x* increasing from 0 to 2.46%. The intense and narrow LO peak, is indicative of the long-range order of the *InP*-host crystal. Figs. 4(e) shows the *Bi* composition dependent full width at half maximum (FWHM) of the *InP$_{1-x}$Bi$_x$* samples. With increasing the *Bi* doping level, the FWHM nearly increases linearly from ~3.2 cm$^{-1}$ (at *x*=0.47%) to 5.9 cm$^{-1}$ (at *x*=2.46%). The increase of the FWHM suggests the quality of the InPBi crystal slowly gets away from the perfect Td symmetry with increasing the *Bi*-content.

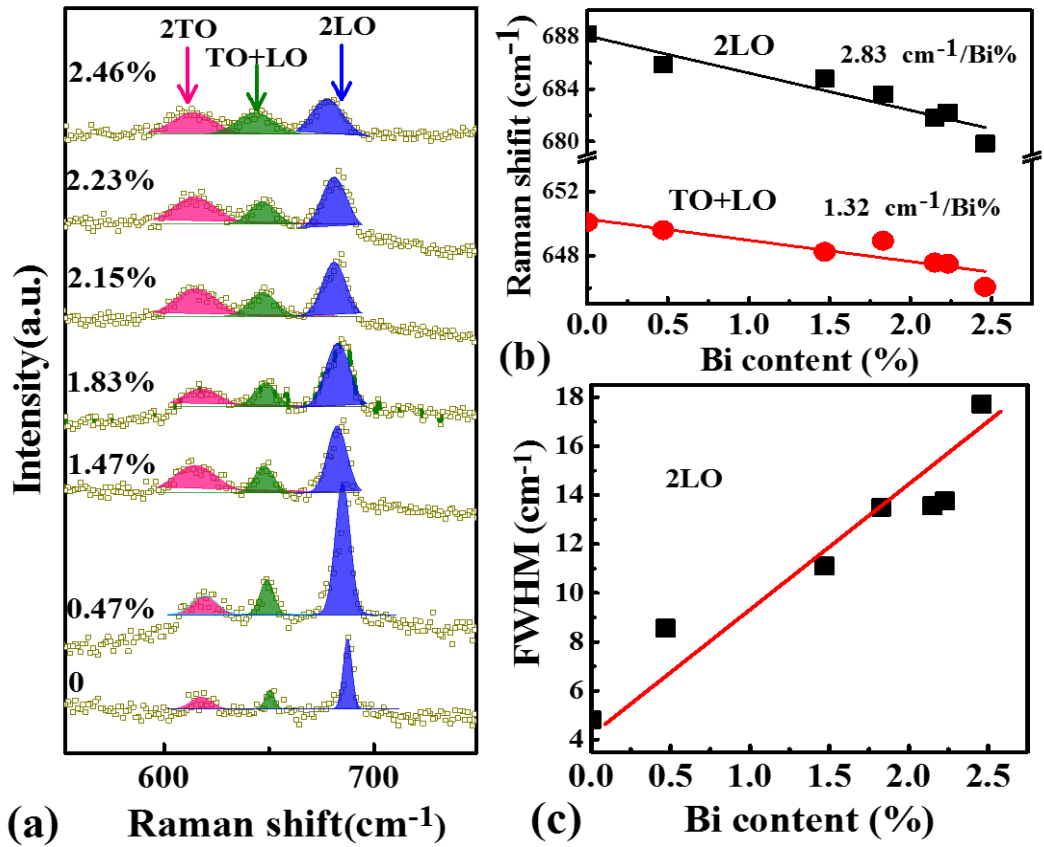

*FIG. 5 (a) Infrared Raman spectra for a series of InPBi samples. The pink sections (2TO), the olive sections (TO+LO) and blue ones (2LO) represent the overtones of TO and LO. (b) The Bi doping level dependent Raman shifts of the 2LO and TO+LO. The dots represent the experimental data, while the straight lines represent the linear fitting progress. (c) The Bi doping level dependent of FWHM of 2LO, where the dots represent the experimental data, and the line guides the eye.*

At last, at the high frequency region (region III), strong second order spectrum between 600 and 700 cm$^{-1}$ were observed. As have been confirmed by other researchers, Raman structures at 615, 649, and 683 cm$^{-1}$ are attributed to a combined creation of LO and TO phonons at various critical points[17], which are the 2TO, TO+LO, and 2LO phonons at Γ point, L point or X point. Raman frequencies of both the 2LO and TO+LO decrease as *Bi*-content increasing, while that of 2TO are insensitive to the *Bi* composition, shown in Figs. 5(b). The Red shifts of the 2LO is around 2.83 cm$^{-1}$ per *Bi*%, which is more than twice of the LO shift. And the red shifts of the TO+LO (1.32 cm$^{-1}$ per *Bi*%) is close to that of the LO. Moreover, the width of the combination band (649 cm$^{-1}$) is smaller than that of the 2TO (615 cm$^{-1}$) and 2LO (683 cm$^{-1}$) overtones indicating that the orderings of the phonon frequencies at Γ, L and X for the TO and LO branches are the reverse of each other[16,17]. Since the second-order LO peak is very sensitive to lattice symmetry, thus it can be used to monitor the degree of disorder in the samples[17-19]. As shown in the Figs. 5(c), the full width at half maximum (FWHM) of 2LO increases almost linearly as *Bi* composition increases, indicating the distortion of the InPBi crystal is increased with increasing the *Bi* doping concentration.

## Conclusions:

In this work, we systematically investigated the vibration properties of the constituent alloys *InP$_{1-x}$Bi$_x$* with $0 \leq x \leq 2.46\%$ by using polarized high resolution micro-Raman. Both *InP*-like and *InBi*-like optical vibration modes (LO) were identified in all the samples, suggesting most of the substitutional Bi-atoms are incorporated into the lattice sites to replace the *P*-atoms. The intensity of the *InBi*-like Raman modes increases exponentially with increasing the Bi doping level. Red-shift of the Raman frequency of the *InP*-like longitudinal optical vibration modes was observed with 1.1 cm$^{-1}$/*Bi*%, while that of *InP*-like optical vibration overtones (2LO) were more than doubled. The crystal distortion of the *InP*-host material is found to increase with increasing the *Bi* doping concentration. In addition, Longitudinal-Optical-Plasmon-coupled (LOPC) modes are identified by comparing the Raman spectra under $Z(X,X)\bar{Z}$ and $Z(X,Y)\bar{Z}$ configurations. The Bi content dependent intensity ratio between the LOPC and *InP*-like LO(Γ) mode has the same trend with the electron concentrations, which can be used as a rapid and sensitive method to determine the relative free carrier density in n-type *InPBi* semiconductors.

## Acknowledgements:

This work was supported by '973 Program' No.2014CB643903, and NSFC Grant